# Sheaf Cohomology and Geometrical Approach to EPR Non-Locality

Menas Kafatos(1) Goro Kato(2), and Sisir Roy(3)

(1) Center for Earth observing
and Space research and School of Computational
Sciences; Physics Department George Mason
University, USA.
E-mail: mkafatos@compton.gmu.edu

(2) Mathematics Department, California Polytechnic
State University, USA.
E-mail: gkato@calpoly.edu

(3) Physics and Applied Mathematics Unit, Indian
Statistical Institute, Calcutta, India
and CEOSR and SCS, George Mason University,
USA.
E-mail: sroy@scs.gmu.edu

## Abstract

A consistent geometrical approach to the EPR non-locality as well as to other non-local effects in quantum mechanics (QM) like the Aharanov-Bohm effect, is possible within the framework of sheaf-cohomology. This approach sheds new light on our understanding on non-local correlations in QM. In general, it also provides a fundamental mathematical approach to foundational problems in physics.

**Keywords :** Sheaf, Cohomology, EPR , Nonlocality

# 1. Introduction:

Non-local correlations present one of the most debated fundamental challenges in quantum mechanics.

Recently, Neeman (1) has attempted a geometrical interpretation of various non-local correlations. It is generally believed that non-local correlations fall into two broad classes : one EPR and the other geometrical. The geometrical aspect is indicated by the Aharonov-Bohm effect (2), Berry phase (3) and other global solutions to quantum gauge theories especially the Yang Mills gauge theories (4), Quantum Gravity (5) etc. A geometrical treatment of gauge theories i.e. a fibre bundle approach has been used where quantum non-localities correspond to pure homotopy classes. On the other hand, EPR non-locality seems to be paradoxical from its very inception. Neeman suggested (1) that perhaps the non-local correlations of EPR can also be derived from geometrical considerations, like all other non-local fibre bundle embedding of other quantum processes so that both kinds of non-localities in QM can be described within the same geometrical framework. However, even though the complementarity between the two types of non-local quantum effects can be grasped within the same framework i.e. in the cases of non-local quantum effects like the Aharanov-Bohm effects and other quantum gauge theories, the homotopical loops do not permit non-local interaction along the paths, whereas in the case of EPR, it requires a measurement at some point along the path.

Several experiments (6) have been performed with pairs of photons which show correlative behaviour over large distances. These may indicate a response in the pairs that are superluminal or even instantaneous correlations (7). Recent developments in sheaf cohomological approach to QM as in (8) shed new light on the geometrical interpretation of both types of non-local correlations in QM. In this paper we shall show how EPR non-locality as well as the measurement i.e. collapse of the state vector can be described with sheaf cohomology. For convenience, we shall briefly describe the main ideas of sheaf cohomology in section 2. In section 3 we shall describe these two types of non-localities within this framework of sheaf cohomology.

## 2. Sheaf Cohomology:

Category theory was created by Eilenberg and MacLane (9) in the forties. It provides powerful and very general methods in various fields of mathematics, especially in algebraic geometry and algebraic analysis.

A category $C$ consists of objects and morphisms. For each pair of objects $X$ and $Y$ in C, we have a set of morphisms from $X$ to $Y$. For morphisms $f$ from $X$ to $Y$ and $g$ from $Y$ to $Z$, the composition of $g$ and $f$ is defined and the composition is a morphism from $X$ to $Z$, which satisfies the associative law: $h(gf) = (hg)f$. For each object $X$, there is a morphism, called the identity morphism $1d_X$, from $X$ to $X$ itself, satisfying $f \circ 1d_X = f$ and $1d_Y \circ g = g$ for any morphism $f$ from $X$ to $Y$ and any morphism $g$ from $Z$ to $X$.

Examples of categories are as follows:

The category of sets consists of objects, the sets and morphisms, the set theoretic maps. The category of abelian groups consists of abelian groups as its objects; and group homomorphisms as its morphisms etc.

An important concept is that of a functor. A (covariant) functor $F$ from a category $C$ to a category $C'$ is defined as follows: for each object $X$ in $C$, $F$ assigns an object $FX$ in $C'$ such that for each morphism $f : X \to Y$ in $C$, assigns a morphism $Ff : FX \to FY$ in $C'$. Then $F$ must satisfy $Fid_x = id_{FX}$ and $F(g \circ f) = Fg \circ Ff$.
 A contravariant functor is a covariant functor from the dual category $C^{opp}$ of $C$ to $C'$, where the dual category consists of the same objects as C and a morphism $f : X \to Y$ in $C^{opp}$ is the morphism $f : X \leftarrow Y$ in $C$. An important example is a presheaf, which will be described in what will follow. Notice that $Hom_C(-,Y)$ and $Hom_C(X,-)$ are examples of a contravariant functor and a covariant functor from $C$ to the category of sets, respectively. The concept of a sheaf was discovered by Leray and Oka(10) in the forties. It was H. Cartan who recognized the equivalence of their notions and developed the theory of cohomology of sheaves in his famous seminars in the fifties, obtaining well-known Theorems in the theory of analytic functions of several complex variables. Further developments were accomplished by Grothendieck(11) in algebraic geometry. The sheaf (cohomology) theory has been employed as a bridge from local information to global information. The following definitions for this implication from local properties to global properties is used.

Let $T$ be a topological space, e.g., Euclidean $n$-space. A presheaf $F$ is an assignment: for every open subset $V$ of $T$

$$V \ ? \ FV$$

where $FV$ is an object in a category satisfying the following axioms. In order to be precise, let T be the category whose objects are open subsets of $T$ and morphisms are only inclusion maps. That is, let $V$ and $V'$ be objects of T, i.e., $V$ and $V'$ are open subsets of $T$.
Then define the set of morphisms from $V$ to $V'$ by

$$\text{Hom}_T(V, V') = \emptyset \text{ for } V \not\subset V'$$
$$\text{and}$$
$$\text{Hom}_T(V, V') = \{ \text{ the inclusion map from } V \text{ to } V' \}$$

$$\text{if } V \subset V'.$$

Let $C$ be a category, e.g., the category of abelian groups. Then, by definition, a *presheaf* is a *contravariant functor* from T to $C$. Namely, $F$ satisfies the following axioms:

**(Presheaf)** For in T, i.e., in $T$, there is a morphism called the restriction morphism $r_V^{v'} : FV \to FV'$ in $C$ such that for $V = V'$, $r_V^{v'}$ is an identity morphism, and for $W \subset V \subset U$, $?_w^V \circ ?_v^U = ?_w^U$ holds. Namely, $F$ is an object of the category

$$\hat{T} = C^{T^{opp}}$$

of contravariant functors from T to $C$, which we write as $F \in C^{T^{opp}}$

Notice that the functor *F* reverses the direction of the arrow from left to right in T to the direction from right to left in *C*, which is the contravariantness of *F*. For example, the assignment $F \to CV$, which is also written as *C(V)* instead of *CV,* where *CV* is the set of all continuous functions defined on a subset *V* of $R^n$. One may obtain many examples by replacing *CV* by the sets of analytic functions, bounded functions, etc. When a presheaf is given, one can ask whether it is possible to obtain a global information from a collection of local data, by "pasting" those local data. The answer is "Yes" if the presheaf further satisfies the following axiom:

**(Sheaf Axiom)** Let *V* be an open set of a topological space *T*, and let { $V_i$ } be an open covering of *V*, i.e., *V* is the union of $V_i$. If $f_i$ and $f_j$ are elements of $FV_i$ and $FV_j$, respectively, satisfying : $f_i$ restricted to the intersection of $V_i$ and $V_j$ equals $f_j$ restricted to the same intersection of $V_i$ and $V_j$, then there exists a unique *f* in *FV* such that the restriction of *f* to each $V_i$ coincides with $f_i$ on $V_i$.

This means that if a presheaf *F* is actually a sheaf, then not only the discrete information data {$F(V_i)$} can be obtained for each covering, but also global information can be obtained by gluing the local data. Then, each local information on $V_i$ is nothing but the restriction of the globally obtained information. Let $\tilde{T}$ be the category of sheaves. That is, $\tilde{T}$ is a subcategory of $\hat{T} = C^{T^{opp}}$. Note that the presheaf of continuous functions is a sheaf. The presheaf of bounded functions is, however, not a sheaf since a locally bounded function like $y = x$ on each open interval in the real line is not bounded on the entire real line. The role of "local existence" versus "global existence" is particularly important in the discussion of time in modern physics. Several authors(12) argue for a solution to the problem of time based on its local existence, while rejecting its possible global existence.
Finally, for the construction of the associated sheaf to a presheaf, see (13). This process is called the sheafification of a presheaf.

## 3. EPR Non-locality and Sheaf Cohomology :

Two fundamental principles, namely the concept of reality and the concept of locality were assumed in the paper of Einstein-Podolsky and Rosen(14). However, these are external asuumptions not integral to quantum mechanics and although philosophically important , they are beyond QM. In order to assert that the physical reality underlying quantum description is incomplete, they considered an idealized physical system consisting of at least two spatially separated subsystems and the quantum state of the system entailing correlation of physical quantities such as polarization of photons or spin of electrons. This state is known as an entangled state. Since, no information is supposed to be propagated from one system to the other far apart at speeds greater than the speed of light, no violation of causality occurs. Bell (15) proved an important theorem that no local realistic theory which satisfies both locality and reality conditions can agree with the predictions of quantum theory. All the experiments so far performed confirm the predictions of quantum theory and violate Bell's inequality (as followed from his theorem). Of course loopholes in the detectors in these experiments need to be closed before a conclusive statement regarding the violation of Bell's inequality is made. It is now believed that some kind of non-locality exists in the quantum framework to explain EPR like situations. This kind of non-local correlation is not directly connected to the propagation of information and hence not physical within the premise of any physical theory or somewhat outside of regular space-time. To understand EPR like situations one needs both the concept of non-locality and the process of measurement in QM i.e. the collapse of the state vector. Within Neeman's geometrical approach, non-local correlations can be explained nicely with the help of homotopy and closed paths but not the local interaction, i.e. measurement. If we consider our framework of sheaf cohomology, both the aspects can be described as follows :

We define $\hat{T}$ as a category of presheaves on the category $T$ associated with a topological space T with values in a product category $\prod_{a \in \Gamma} C_a$. More precisely, $\hat{T}$ is the category of contravariant functors from the category T associated with a topological space $T$ to a product category $\prod_{a \in \Gamma} C_a$ of categories where $\Gamma$ is an index set. The category T is said to be the *generalized time space* (or *generalized time category*) when the real line R is embeddable in T . Namely,

$$\hat{T} = \left( \prod_{a \in \Gamma} C_a \right)^{T^{opp}}$$

To be more explicit, for an object $V$ in $T$, i.e., an open set $V$ of T, and for an object $P$ in $\hat{T}$, we have $P(V) = (P_a(V)), a \in \Gamma$, where each $P_a(V)$ is an object of $C_a$. Recall that an *entity* is a presheaf $P$ in $\hat{T}$ where $\{C_a, a \in \Gamma\}$ represents the totality of physical categories of all entities. Note that $\hat{T}$ represents matter like elementary particles, atoms, molecules etc. $C_o, C_1, C_2$ represent generalized time category, micro world and macro world respectively. $C_1, C_2$ are considered as discrete categories with structures.

Let e and e' be the elementary quantum subsystems as electrons used in EPR situation as objects of i.e. each electron is an entity. Let e and e' be the associated presheaves. When e and e' are correlated, by the definition, then the pair (e, e') becomes an entity which we denote by e̲ = ( e, e'). That is, for a generalized time period $V$, let e̲ $(V) = $ (e , e') $(V)$ $= ($ e$(V),$ e' $(V))$ be an object in an entangled state in $C_1$ or $C_2$.

Therefore, the state of e̲ in $C_2$ is determined by $V$ alone. That is, e̲$(V) = $ (e $(V),$ e' $(V))$ is independent of the positions in $C_2$. In particular the states of e and e' are determined only by a generalized time period $V$.

Now this kind of communication happens over generalized time and we do not have any real propagation over the physical category. So non-locality in EPR like situations have a consistent explanation in the premise of sheaf cohomology which is much broader framework for the description of underlying reality.

Let us discuss now the problem of measurement which is also one of the fundamental aspects in EPR situations. Let $P$ and $Q$ be objects of $\hat{T}$, i.e., the category of presheaves over $T$.
When there is a morphism from $P(U)$ to $Q(U)$ in a non-discrete category, where measurement takes place, then $P$ is said to be measurable, or observable by $Q$ over the generalized time period $U$. In this context, $Q$ is the observer and $P$ is the observed system. That is, $P$ is measurable during the generalized time period $U$ by $Q$ when there exists a natural transformation from $P$ to $Q$ over the specified $U$. In the usual sense of measurement theory, there is a real time interval over which measurement takes place. In general, an object $P$ in $\hat{T}$ is evaluated at a generalized time period $U$, $P(U)$ is manifested, hence inducing the usual linear time *t*. In our formalism, the specific time *t* is induced by the associated presheaf $\tau$ evaluated at a generalized time period. This means that *t* is presheafified satisfying t=$\tau$ (V) for some object *V* in *T* see (8) for presheafification of the notion of time . Let us now return to the above EPR like situation for entangled two electrons:

When e is observed by *P* over a generalized time period *V*, there is morphism form e*(V)* to *P(V)*. Since e and e′ are entangled, for this *V*, an morphism from e′*(V) to* e*(V)* induces a morphism from e′*(V)* to *P(V)* as well. Therefore, there is morphism from the pair e*(V)=(*e*(V),* e′*(V))* to *P(V)*. So by conducting measurement on one we can effectively get information for the entire system regardless of the distance between the two electrons as they never cease to be entangled. This kind of correlation between two electrons or entities in the micro-domain $C_1$ is based on the fact that a pair forms a new undivided entity over generalized time *V*. This will be discussed in detail in subsequent papers.